\documentclass[letterpaper]{article} 
\usepackage[]{aaai2026}  
\usepackage{times}  
\usepackage{helvet}  
\usepackage{courier}  
\usepackage[hyphens]{url}  
\usepackage{graphicx} 
\usepackage{natbib}  
\usepackage{caption} 
\usepackage{graphicx}
\usepackage{amsmath} 
\usepackage{algorithm}
\usepackage{algorithmic}

\usepackage{array}
\usepackage{newfloat}
\usepackage{listings}
\DeclareCaptionStyle{ruled}{labelfont=normalfont,labelsep=colon,strut=off} 
\floatstyle{ruled}
\newfloat{listing}{tb}{lst}{}
\floatname{listing}{Listing}
\title{QuantumChem-200K: A Large-Scale Open Organic Molecular Dataset for Quantum-Chemistry Property Screening and Language Model Benchmarking}
\author{
    Yinqi Zeng\textsuperscript{\rm 2}\equalcontrib,
    Renjie Li\textsuperscript{\rm 1,2}\equalcontrib
}
\affiliations{
    \textsuperscript{\rm 1} Holonyak Micro and Nanotechnology Laboratory, University of Illinois Urbana-Champaign\\
    \textsuperscript{\rm 2} Electrical and Computer Engineering, University of Illinois Urbana-Champaign\\
    renjie2@illinois.edu
}

\begin{document}

\maketitle

\begin{abstract}
The discovery of next-generation photoinitiators for two-photon polymerization (TPP) is hindered by the absence of large, open datasets containing the quantum-chemical and photophysical properties required to model photodissociation and excited-state behavior. Existing molecular datasets typically provide only basic physicochemical descriptors and therefore cannot support data-driven screening or AI-assisted design of photoinitiators. To address this gap, we introduce QuantumChem-200K, a large-scale dataset of over 200,000 organic molecules annotated with eleven quantum-chemical properties, including two-photon absorption (TPA) cross sections, TPA spectral ranges, singlet–triplet intersystem crossing (ISC) energies, toxicity and synthetic accessibility scores, hydrophilicity, solubility, boiling point, molecular weight, and aromaticity. These values are computed using a hybrid workflow that integrates density function theory (DFT), semi-empirical excited-state methods, atomistic quantum solvers, and neural-network predictors.
Using QuantumChem-200K, we fine-tune the open-source Qwen-2.5-32B large language model to create a chemistry AI assistant capable of forward property prediction from SMILES. Benchmarking on 3000 unseen molecules from VQM24 and ZINC20 demonstrates that domain-specific fine-tuning significantly improves accuracy over GPT-4o, Llama-3.1-70B, and the base Qwen2.5-32B model, particularly for TPA and ISC predictions central to photoinitiator design. QuantumChem-200K and the corresponding AI assistant together provide the first scalable platform for high-throughput, LLM-driven photoinitiator screening and accelerated discovery of photo-sensitive materials.
\end{abstract}

\section{Introduction}

Recent advances in large language models (LLMs) have demonstrated strong capabilities in general reasoning, code generation, and text summarization \cite{yang2024harnessing, kumar2024largelanguagemodels, patil2024review, zhao2023survey}. However, despite this rapid progress, LLMs still face limitations in highly specialized scientific domains that demand deep technical knowledge and structured data, with materials science as an example. Many general-purpose chemical datasets—such as GDB, QM7, QM9, and ZINC20—have been proposed and updated in the past decades \cite{hoja2021qm7x, pinheiro2020machine, irwin2020zinc20, letovsky1998gdb}. While neural-network–based approaches have seen used extensively in drug and protein discovery in recent years \cite{wang2018computational, ali2024deepep}, a huge blank space remains in extending these methods to broader materials discovery. Progress in this area increasingly relies on high-quality, domain-specific datasets that link molecular structure to quantum-chemical behavior. Such datasets are scarce, and existing public collections seldom include the photo-chemical and quantum features needed for specific materials discovery tasks.

\begin{figure}[htb!]
  \centering
  \includegraphics[width=0.75\linewidth]{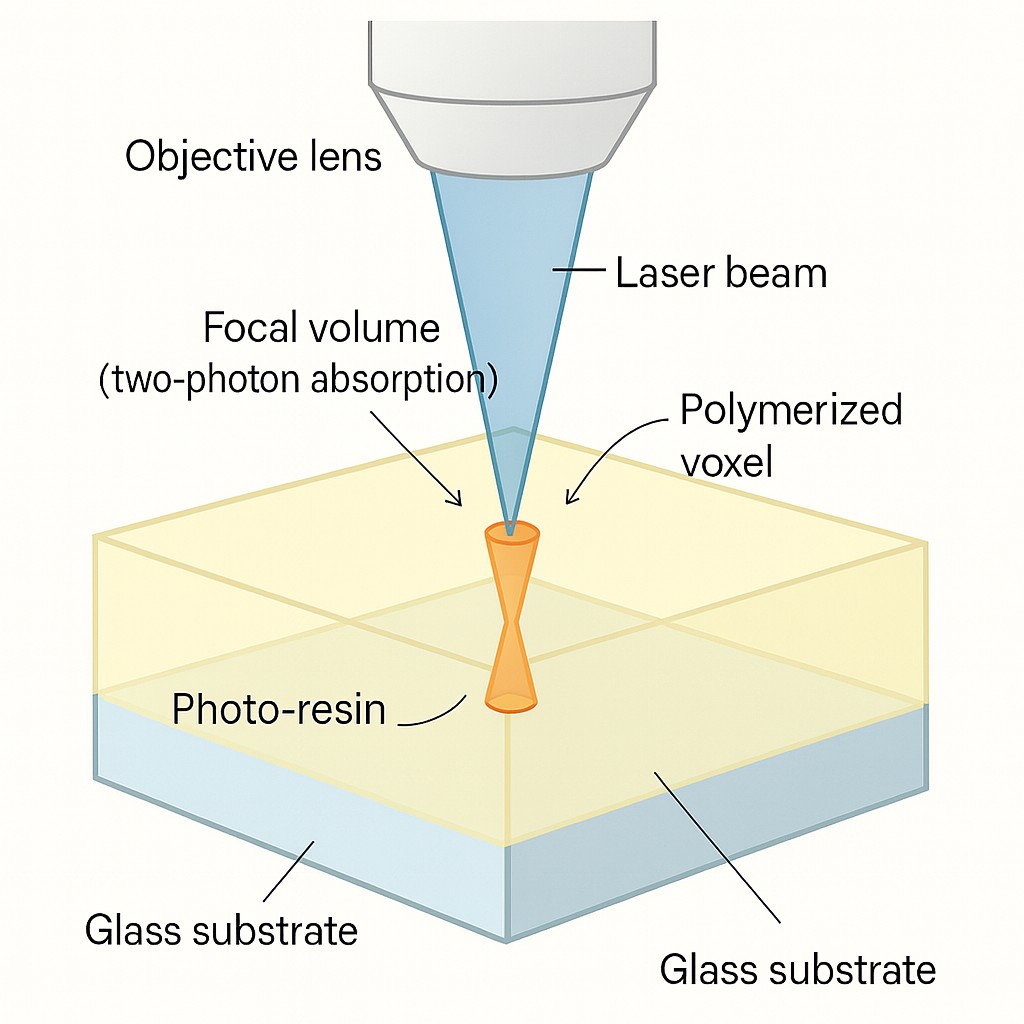}
  \caption{Schematic of DLW: A focused photon beam is confined to a single voxel within the photo-resin. At this focal point, the photoinitiator absorbs light, undergoes dissociation, and generates reactive radicals. These radicals initiate polymerization with nearby monomers in the resin, enabling the formation of solid structures with nanoscale precision.}
\end{figure}

Photodissociation is a process of a photoinitiator molecule breaking into radical fragments after absorption of photons. It is a central mechanism during photopolymerization. Figure 1 shows the specific mechanism of  direct laser writing (DLW) which involves polymerization. Norrish Type I photoinitiators \cite{jagtap2022selfinitiated, rutsch1996recent} work by absorbing either one photon or two infrared photons (typically around 780 nm). This excitation moves the molecule from the ground state ($S_0$) to the excited singlet state ($S_1$). It then quickly undergoes intersystem crossing to the triplet state ($T_1$), where alpha-cleavage (C–C bond breaking) occurs and the molecule splits into two reactive radicals. These radicals immediately react with the surrounding liquid monomers and initiate polymerization, turning the resin into solid material. This mechanism enables two-photon polymerization (TPP), which is widely used in bioprinting, sub-micron additive manufacturing, and subsurface-DLW. The use of near-infrared light provides strong spatial confinement, low scattering, and allows precise 3D structuring at the nanoscale \cite{lee2006recent, nguyen2017tpp, zhou2015tppaccuracy}.

\begin{figure*}[htb!]
  \centering
  \includegraphics[width=0.85\linewidth]{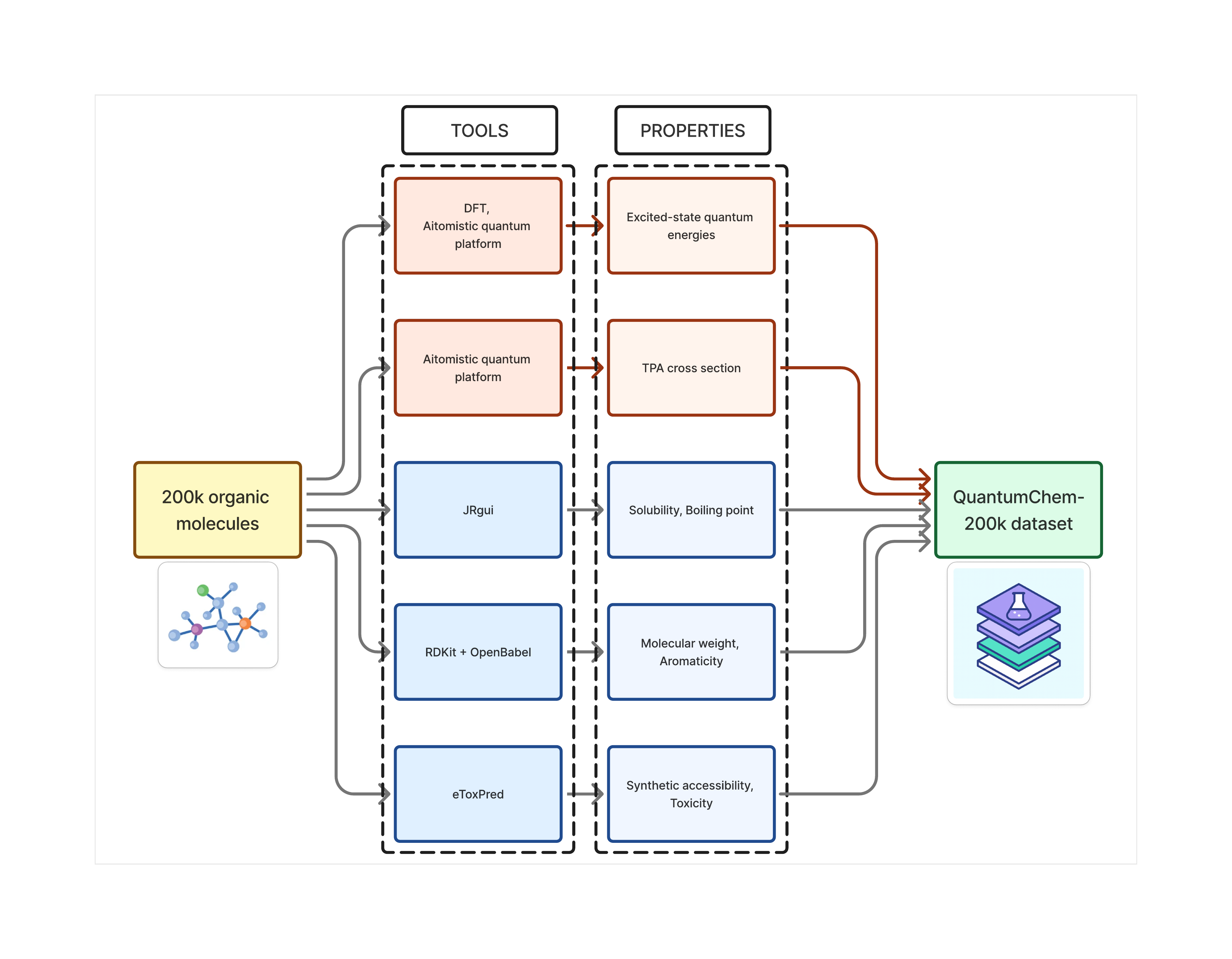}
  \caption{General workflow for the QuantumChem-200K dataset curation}
\end{figure*}

Despite the industrial and scientific importance of photoinitiators, progress in discovering new candidates has been slow, largely because commercial photoresin formulations are usually proprietary, limiting transparency into their compositions. Moreover, there is no large-scale dataset currently that provides collection of photodissociation-relevant quantum properties required for data-driven screening. Moreover, it is difficult to predict photodissociation quantum yield computationally; experimental determination is also labor-intensive and molecule-specific \cite{magnotta1980photodissociation, ye2002photodissociation}. As a result, the field lacks the data foundation needed for machine-learning-based discovery, LLM-assisted molecular design, or high-throughput photoinitiator screening.

In this work, we address this gap by curating an open dataset of over 200,000 organic molecules containing 11 key properties closely linked to photodissociation efficiency and photoinitiator performance. Figure 2 and Table 1 have listed all the properties computed and their description. These properties include two-photon absorption (TPA) cross sections, maximum TPA strength and absorption ranges, $S_1$ to $T_1$ state intersystem crossing (ISC) energies, fundamental thermodynamic descriptors (boiling point, molecular weight), molecular hydrophobicity (logP), synthetic accessibility, aromaticity, and toxicity metrics. Each property was selected based on mechanistic relevance: efficient TPP requires large TPA cross sections at ~780 nm, rapid population transfer from $S_1$ to $T_1$ requires small ISC energetic gaps, and practical candidates must also be safe, synthetically accessible, and compatible with common photoresin formulations. 

Beyond dataset construction, we fine-tuned the open-source Qwen-2.5-32B LLM with QuantumChem-200K to develop a chemistry AI assistant capable of forward property prediction for arbitrary input SMILES strings. The AI assistant is evaluated on an external benchmark composed of two independent molecular datasets (VQM24 and ZINC20), using a weighted mean absolute error (wMAE) metric tailored for chemistry/materials tasks. Futhermore, we have compared the performance of prediction accuracy between our fine-tuned AI assistant with other common LLMs: Qwen2.5-32B, GPT-4o, Llama-3.1-70B. This evaluation demonstrates that LLM-based property prediction can be significantly enhanced when the model underwent domain-specific large-scale quantum-chemical data fine-tuning.

Overall, this work provides 
\begin{itemize}
    \item The first large-scale, open photoinitiator-focused molecular dataset for training and evaluating AI assistants and agents for scientific research in materials and chemistry.
    \item A fine-tuned chemistry AI assistant capable of predicting 11  essential properties of photoinitiators.
    \item A scalable and high-throughput screening pipeline for evaluating new photoinitiators generated by AI assistants for two-photon polymerization  applications.
\end{itemize}

\begin{table*}[htb!]
\small
\centering
\begin{tabular}{|p{5cm}|p{6cm}|p{4cm}|}
\hline
\textbf{Property}  & \textbf{Description} & \textbf{Curation Tool}\\ \hline

TPA cross section ($\sigma$)$\uparrow$ &Molecule’s probability of simultaneously absorbing two photons, measured in GM units&  Aitomistic platform \cite{Dral2024MLatom3}\\

Wavelength range of absorption $\uparrow$ & The wavelength range where a molecule absorbs photons with cross section over 20GM &  Aitomistic platform  \\

Intersystem crossing energy (ISC) $\downarrow$ & Quantum energy gap between the first excited singlet state ($S_1$) and triplet state ($T_1$) & Aitomistic and DFT \cite{thiel2014semiempirical}\\

Toxicity $\downarrow$ and Synthetic accessibility $\uparrow$& The potential biological harm; Easiness of synthesizing the molecule &  eToxPred NN \cite{Pu2019eToxPred}\\

Solubility, Boiling point and hydrophilicity $\uparrow$ & ug/mol ; $^\circ$C ; unitless &  JRgui \cite{Shi2017JRgui}\\

Molecular weight and aromaticity $\downarrow$ & g/mol; Whether aromatic rings exist in the molecule &  openBabel \cite{o2011open}, RDKit\\
\hline
\end{tabular}
\caption{Description of the photochemical and quantum-relevant molecular properties computed for the QuantumChem-200K dataset.
The arrows in the first column indicate the preferred values (↑ = larger value favored; ↓ = smaller value favored). The second column defines each property and, where appropriate, specifies its physical units. The third column lists the quantum computational tools, neural network (NN) models, or computing platforms used to generate the corresponding values.}
\label{tab:llm4pcsel_results}
\end{table*}

\section{Related Works}

Large molecular datasets have become central to data-driven chemistry, providing broad chemical coverage across small organics, drug-like compounds, and functional materials. Some general moleuclar database include QM7, QM9, GDB-17, ZINC20, and the Open Molecule Genome (OMG) \cite{hoja2021qm7x,letovsky1998gdb,irwin2020zinc20,ramakrishnan2014quantum} which typically contains SMILES/InChI identifiers and basic physicochemical descriptors such as logP, molecular weight, and functional-group counts. In parallel, a variety of domain-specific datasets have been developed for specialized areas of chemistry and materials science. For example, dataset curated focusing on inorganic and solid-state materials, organic optoelectronic or nanomaterial systems \cite{zakutayev2018open, joung2020experimental, yan2020nanomaterial, jeliazkova2015enanomapper}. Such datasets extend general molecular collections into specific material classes and support applications that require targeted chemical or structural representations.

Recent advances in time-dependent DFT (TD-DFT) and large-scale HPC clusters allowed the curation of large quantum-chemistry datasets that include optimized geometries and quantum-electronic properties. Beyond QM7/QM9, datasets such as ANI-1ccx, VQM24, and PubChemQC provide DFT-level energies, excitation properties, and conformer landscapes \cite{smith2020ani1ccx, khan2024vqm24, nakata2017pubchemqc}. These quantum-optimized structures are essential for downstream computations that cannot be inferred from SMILES or graphs. For example, properties including the TPA spectra, ISC energy, and excited-state energetics, which are crucial for photoinitiator and photochemical-material design, need both optimized geometry and TD-DFT to compute. 

Machine-learning models for molecular property prediction have traditionally been dominated by graph neural networks (GNNs), particularly those based on message passing neural networks (MPNNs). Within quantitative structure–property relationship studies, GNNs operate directly on molecular graphs: atoms are represented as nodes, bonds as edges are used to update node embeddings through iterative message passing \cite{hirschfeld2020uncertainty, tang2023mpnn_property_prediction, taskinen2003prediction}. This framework allows GNNs to learn chemically meaningful patterns through molecular topology, hence improving the prediction accuracy over most properties. By contrast, there is fewer research based on fine-tuning LLMs for property prediction \cite{jacobs2024regression, liu2024moleculargpt, xiao2024proteingpt} since LLMs are trained on text-based representations such as SMILES. As a result, for broad molecular property prediction like boiling point, solubility, logP, and toxicity, GNNs are typically better than LLMs, especially when extrapolating outside the distribution of the training sets. This gap has motivated recent hybrid LLM–GNN architectures that combine linguistic priors from LLMs with the structural precision of GNNs to improve property-prediction accuracy \cite{li2025hybrid}.

However, when the target properties rely on electronic-structure changes, such as excited-state energies, transition dipoles, or intersystem crossing energy, LLM-based approaches offer significant advantages. When fine-tuned on high-quality quantum chemistry data, LLMs can approximate energy landscapes and excitation trends that depend on subtle global context rather than strictly local graph topology. Moreover, LLMs integrate naturally with LangChain-based autonomous agentic pipelines, enabling seamless incorporation of literature retrieval, multi-step reasoning (chain-of-thought), self-reflection, persistence, long-token text interpretation, and tool calling. This flexibility and portability make fine-tuned LLMs particularly suitable for workflows that combine molecular prediction, knowledge extraction, and automated scientific discovery.

\begin{table}[htb!]
\small
\centering
\begin{tabular}{|p{2.8cm}|p{2cm}|p{2.3cm}|}
\hline
\textbf{Traits}  & \textbf{134k} & \textbf{77k}\\ \hline
Length of SMILES & Up to 9 heavy atom &  Up to 25 heavy atoms \\
Chemical elements & CHNOF &  CHNOFSBrClSi  \\
Properties computed & All & All Except ISC \\

\hline
\end{tabular}
\caption{Comparison between source 134K and 77K datasets}
\label{tab:llm4pcsel_results}
\end{table}
\section{QuantumChem-200K Dataset}

The QuantumChem-200K molecular dataset is generated using a multi-stage pipeline (Figure 2) integrating density functional theory (DFT), semi-empirical quantum chemistry tools, and neural-network–based predictors. 210K source SMILES data are taken from two open molecular databases: a 77K dataset and a 134K dataset (Table 2).

\subsection{Source Datasets}
77k Monomer Subset (OMG Dataset):
the first component of the dataset consists of 77k monomers collected from the Open Macromolecular Genome (OMG), a large library of synthetically accessible monomer and polymer structures \cite{Kim2023OMG}. OMG provides molecular structures containing up to 25 heavy atoms (C, H, N, O, F, Br, Cl, Si, P, S). This broader chemical space naturally includes diverse chromophores, electron-donating and electron-withdrawing motifs, and bond-cleavable groups that are chemically relevant to photoinitiation and photodissociation. Many known Norrish-type I photoinitiators fall within this dataset, making OMG an effective starting point for large-scale candidate generation.

134k Organic Molecule Subset:
the second component consists of 134k small organic molecules extracted from the curated dataset QM9 \cite{ramakrishnan2014quantum}. This collection includes molecules composed of CHNOF elements with up to 9 heavy atoms. The dataset provides optimized ground-state geometries, which we further leveraged to compute singlet and triplet excited-state energies. Excited-state computations were performed using MNDO (Modified Neglect of Diatomic Overlap) \cite{thiel2014semiempirical}, a semi-empirical electronic structure method suitable for CHNOF molecules. MNDO is used here to efficiently evaluate vertical excitation energies across multiple excited-state, enabling systematic extraction of singlet ($S_1$) and triplet ($T_1$) energy levels for ISC estimation.

Both sub-datasets naturally contain chemical structures resembling experimentally validated photoinitiators and many molecules display desirable features such as large TPA cross-sections, favorable $S_1$ to $T_1$ intersystem energy gaps, and high quantum yield. This confirmed the suitability of combining the two chemical databases for constructing the QuantumChem-200K dataset. Details of the contents of each sub-dataset are summarized in Table 2.

\subsection{Property Computation}
Calculations of the 11 quantum-chemical properties are introduced here, per the content of Table 1. 
\subsubsection{Intersystem Crossing (ISC) Energy}

To quantify the \( S_1 \rightarrow T_1 \) ISC energy, we developed a hybrid DFT–semi-empirical workflow:

\begin{enumerate}
    \item \textbf{Geometry optimization} was carried out using the B3LYP functional with the 6-31G(2df,p) basis set, providing reliable ground-state geometries for subsequent property calculations.
    \item \textbf{Excited-state energies} for both singlet and triplet manifolds were computed using MNDO(ODM2*) hosted on Aitomistic(AIQM1), which provides efficient semi-empirical estimates for CHNO systems.
\end{enumerate}

The ISC energy is defined as:
\[
\Delta E_{\mathrm{ISC}} = E_{S_1} - E_{T_1},
\]
where smaller gaps generally correspond to faster ISC rates and improved photodissociation efficiency.

\subsubsection{Two-Photon Absorption (TPA) Calculation}

Two-photon absorption (TPA) cross-sections $\sigma$ was computed using the Atomicistic Quantum Platform, which evaluates dynamic two-photon responses across the \(600\text{–}850\,\mathrm{nm}\) spectral range. All computations assume Et55.4 (1-octanol) as the solvent, consistent with common photoinitiator characterization conditions.

For each molecule, we record:
\begin{itemize}
    \item the \textbf{maximum TPA cross-section} across the spectrum,
    \item the \textbf{TPA absorption window},
    \item the \textbf{specific TPA cross-section at \(780\,\mathrm{nm}\)}, a standard near-infrared excitation wavelength in two-photon polymerization.
\end{itemize}

\subsubsection{Toxicity and Synthetic Accessibility}

Toxicity (Tox) and synthetic accessibility (SA) scores were predicted using \texttt{eToxPred}, a neural network trained on experimentally annotated toxicity datasets and retrosynthetic complexity metrics. Both scores range from 0 to 1, where:
\begin{itemize}
    \item \(0\) indicates low toxicity or easy synthesis,
    \item \(1\) indicates high toxicity or challenging synthesis.
\end{itemize}

\begin{figure*}[htbp!]
  \centering
  \includegraphics[width=0.85\linewidth]{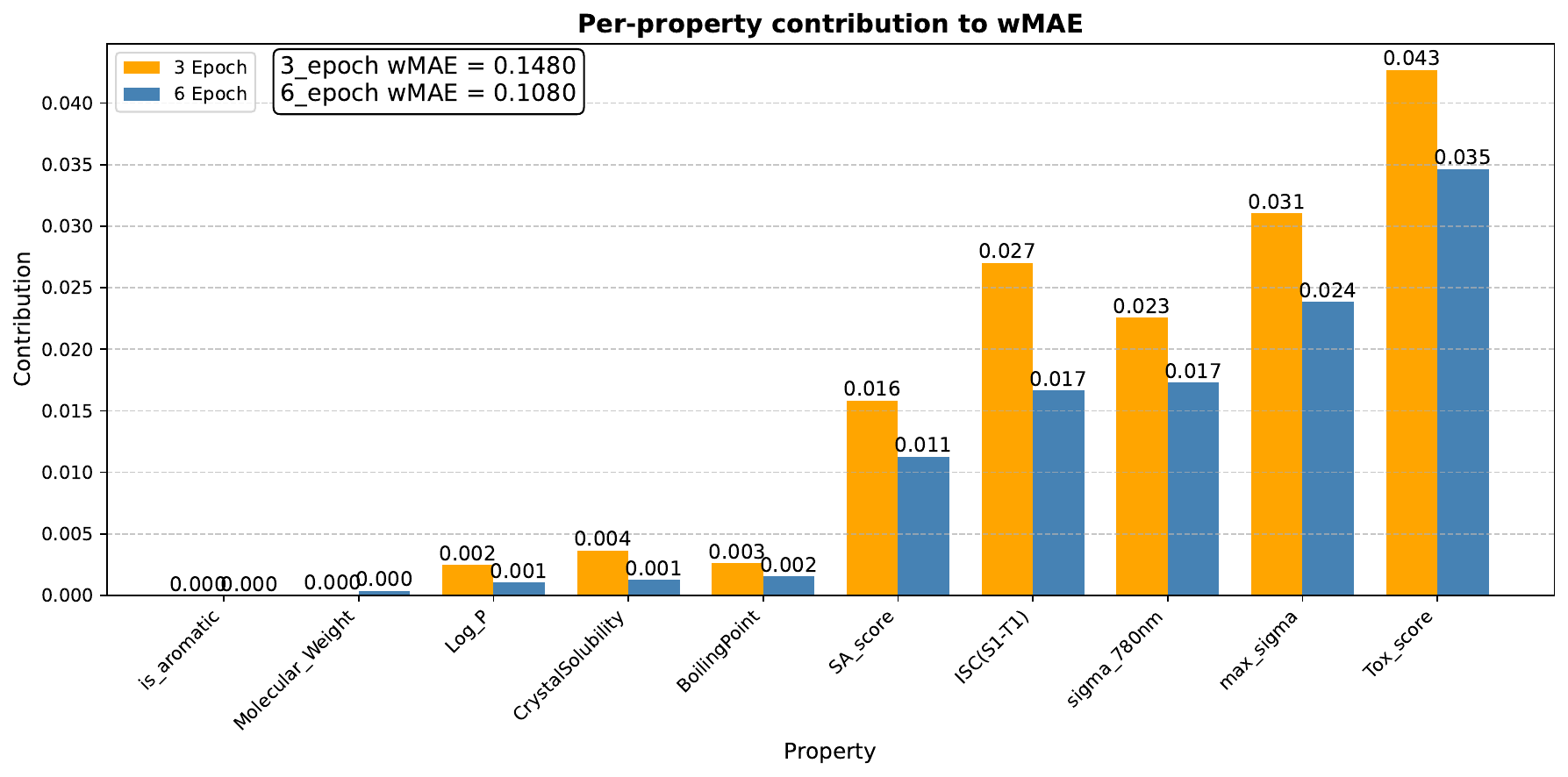}
  \caption{wMAE of the AI assistant (Qwen2.5-32B) for each property during fine-tuning on the QuantumChem-200K dataset, where orange and blue are the wMAE at 3 and 6 epochs of training, respectively. Number on top of each bar is the per-property contribution to the overal wMAE. wMAEs here is calcualted with 100 randomly sampled data points.}
\end{figure*}

\subsubsection{Other Physicochemical Properties}

Additional physicochemical descriptors---including \textbf{boiling point}, \textbf{solubility}, \textbf{Molecular Weight}, \textbf{hydrophilicity (logP)}, and \textbf{aromaticity}---were computed using openBabel, RDKit, and the JRgui Python suite. JRgui extracts functional-group statistics and substructure patterns using cheminformatics toolkits, ensuring that predicted photoinitiators satisfy practical constraints related to handling, solubility in photoresin systems, and chemical compatibility under ambient laboratory conditions.

\section{Results and Discussion}
Our chemistry AI assistant is fine-tuned on the QuantumChem-200K dataset from Qwen2.5-32B base model. A custom weighted MAE (wMAE) metrics is used to evaluate our fine-tuned AI assistant and benchmark the AI assistant's prediction accuracy against GPT-4o, Llama-3.1-70B and Qwen2.5-32B baselines. 

\subsection{wMAE metrices}
To evaluate prediction performance across the ten heterogeneous molecular properties in our dataset, we adopt the weighted Mean Absolute Error (wMAE) metric. Unlike the standard MAE—which treats all properties as equally scaled and equally represented—wMAE introduces a task-dependent reweighting factor that accounts for both numerical range differences and uneven sample availability across property types. This makes wMAE highly suitable for our photoinitiator-oriented dataset, where the physical quantities span different magnitudes and some properties are sparsely represented in the test bank. Our wMAE follows the NeurIPS Open Polymer Prediction 2025 competition definition \cite{Kaggle2025OPP}. 

\begin{equation}
\mathrm{wMAE}
= \frac{1}{\lvert\mathcal{M}\rvert}
  \sum_{M \in \mathcal{M}}
  \sum_{i \in \mathcal{I}(M)}
  w_i \,\left|\,\hat{y}_i(M) - y_i(M)\right|
\end{equation}

\begin{equation}
w_i
= \left(\frac{1}{r_i}\right)
  \left(
    \frac{
      K \sqrt{\frac{1}{n_i}}
    }{
      \displaystyle \sum_{j=1}^{K} \sqrt{\tfrac{1}{n_j}}
    }
  \right)
\end{equation}

\noindent\textbf{Symbols:}

$\mathcal{M}$: set of evaluated molecules/monomers; 

$\mathcal{I}(M)$: index set of properties available for molecule $M$ (missing properties are skipped).

$\hat{y}_i(M)$ / $y_i(M)$: predicted / ground-truth value of property $i$ for $M$.

$w_i$: weighting factor for property $i$.

$r_i=\max(y_i)-\min(y_i)$: observed range of property $i$ (computed on the evaluation split).

$n_i$: number of evaluation samples with a valid $y_i$.

$K$: total number of property prediction tasks included in the metric.

The factor $1/r_i$ rescales properties to reduce the effect of differing units/ranges, while $\frac{K \sqrt{1/n_i}}{\sum_{j=1}^{K}\sqrt{1/n_j}}$ allocates relatively larger weight to rarer properties (smaller $n_i$) and is normalized across tasks before the range scaling is applied.

\subsection{Evaluating the Chemistry AI Assistant} 

We first evaluated the predictive performance of our fine-tuned model at two training checkpoints---after 3 epochs and after 6 epochs---using a set of 100 molecules randomly sampled from the QuantumChem-200K dataset (Figure 3). Across these two checkpoints, the overall wMAE decreased by approximately 30\%, demonstrating that additional fine-tuning substantially improves the model's ability to capture structure--property relationships encoded in the SMILES representations.

While the wMAE distribution reveals a spike in the toxicity (Tox score) as shown in Figure 3, this trend is expected. Toxicity is inherently difficult to infer directly from molecular strings because it depends on subtle structural motifs, rare functional groups, and complex biological interactions that may not be well represented in the training data. In contrast, the model shows consistent and significant improvement on quantum-chemical properties---particularly the TPA cross-section at \(780\,\mathrm{nm}\) (\(\sigma_{780}\)) and the ISC, which correlate strongly with electronic structure patterns that LLMs can progressively internalize during training.

Quantitatively, both \(\sigma_{780}\) and ISC exhibit improvements of roughly 40\%, with the wMAE of \(\sigma_{780}\) decreasing from \(0.027\) to \(0.017\), and the wMAE of ISC decreasing from \(0.023\) to \(0.017\). These trends indicate that the model is increasingly capable to associate SMILES-level structural cues with corresponding photochemical behaviors, which suggests that LLMs, when fine-tuned on sufficiently large quantum-chemical datasets, can progressively approximate the underlying physical mechanisms governing photodissociation and excited-state transitions.

\begin{figure}[htbp!]
  \centering
  \includegraphics[width=0.95\linewidth]{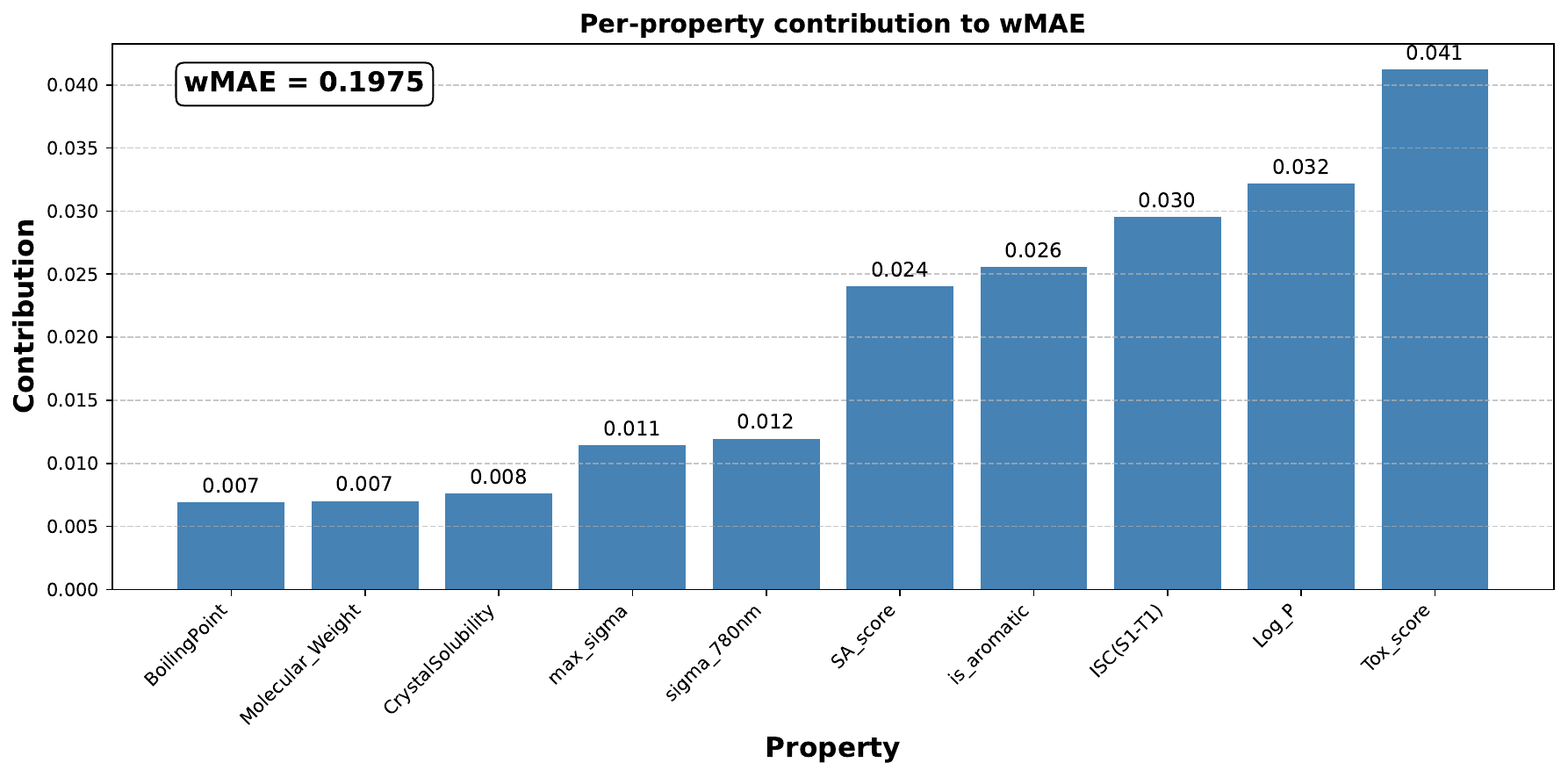}
  \caption{Final AI assistant evaluation with wMAE for the 3000 testbank, showing an overall wMAE of 0.1975.}
\end{figure}

\begin{figure*}[htbp!]
  \centering
  \includegraphics[width=0.85\linewidth]{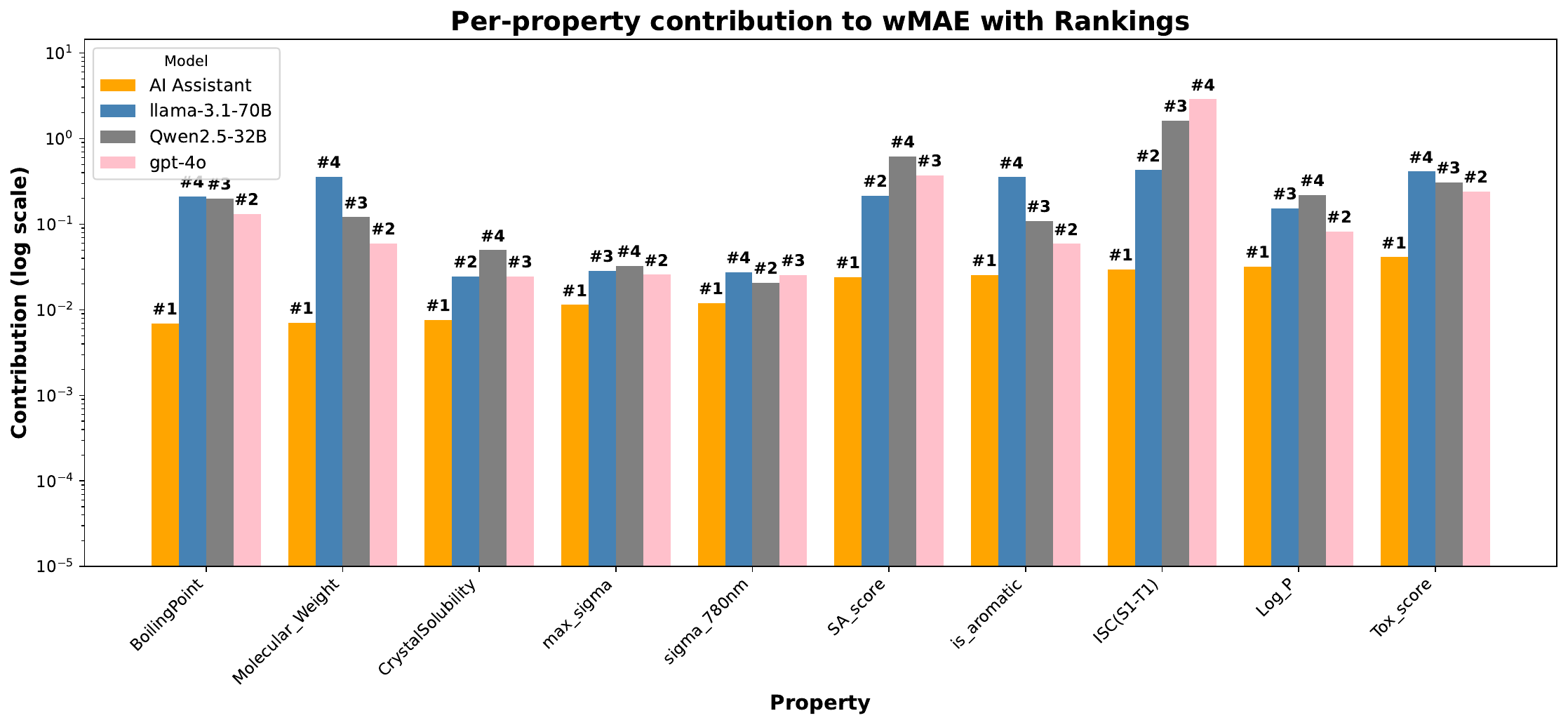}
  \caption{Ranking the wMAE of the AI assistant (orange), llama-3.1-70B (blue), Qwen2.5-32B (gray), and gpt-4o (pink) on the 3000 testbank for each property, where number on top of each bar is the ranking. Overall wMAE value of each model is recorded in Table 3. }
\end{figure*}

\begin{table}[htb!]
\small
\centering
\begin{tabular}{|p{1cm}|p{1.2cm}|p{1.2cm}|p{1.2cm}|p{1.2cm}|}
\hline
\textbf{}  & Fine-tuned AI Assistant & Llama-3.1-70B & Qwen2.5-32B & GPT-4o\\ \hline
wMAE & \textbf{0.1975} & 2.2195 & 3.3038 & 3.9200 \\

\hline
\end{tabular}
\caption{Overall wMAE of different LLMs on the testbank.}
\label{tab:llm4pcsel_results}
\end{table}

\subsection{Testbank Setup}
To rigorously evaluate the generalization ability of our model beyond the training dataset, we constructed a testbank of 3,000 previously unseen molecules collected from two independent sources. We first selected 1,000 molecules from the VQM24 dataset \cite{khan2024vqm24}, which contains more than 836,000 geometry-optimized monomer structures. Only CHNOF molecules were retained to ensure compatibility with our ISC workflow, since MNDO supports vertical excitation energy calculations only for CHNO elements. The second subset consists of 2,000 molecules extracted from ZINC20 \cite{irwin2020zinc20}, a large and chemically diverse database containing drug-like structures with up to 20 heavy atoms (CHNOFBrSiSCl). This subset introduces significantly broader chemical diversity than the training set, including functional groups not present in QuantumChem-200K. These features make ZINC20 an ideal source for testing robustness and out-of-distribution generalization.

After assembling both subsets, we applied the same automated workflow described in Figure 2 to compute all 11 quantum-chemical and physicochemical properties for consistency. This ensures that the testbank provides an unbiased and property-complete benchmark for evaluating the performance of our AI assistant as well as other baseline LLMs.

\subsection{Benchmarking}
Using the full 3,000-molecule testbank, we benchmarked the fine-tuned AI assistant and compared its performance against several state-of-the-art baseline LLMs. Figure 4 presents the wMAE of our fine-tuned model. Compared to the wMAE during fine-tuning (Figure 3), the wMAE on the external testbank is slightly higher, which is expected given the increased structural diversity and the presence of many scaffolds the model has never encountered. As observed previously, toxicity (Tox) exhibits the highest error, which shows the inherent difficulty of predicting biological toxicity from SMILES strings. An increase in logP error is also observed, likely due to the more complex functional groups and longer backbones present in ZINC20.

Despite these challenges, the model performs strongly on photophysics-relevant tasks. The predictions for TPA $\sigma$ at 780 nm and the maximum $\sigma$ value remain highly accurate, with wMAE values of 0.011 and 0.012, respectively as in Figure 4. These results indicate that the model has learned robust structure–property relationships governing nonlinear optical behavior. Quantum-chemical property prediction, such as the ISC,  remains competitive as well, with only a modest drop in accuracy due to the inclusion of non-CHNOF structures that lie outside the distribution used during training.

Table 3 reports the overall wMAE across all tested LLMs, and Figure 5 visualizes the per-property contribution to MAE (log-scaled for readability) with rankings. Our fine-tuned model shows consistently superior performance across all properties. In particular, it achieves much higher ISC prediction accuracy than the three baseline LLMs. Among the baselines, Llama-3.1-70B performs best, while GPT-4o performs worst. Importantly, all three untrained LLMs exhibit wMAE values more than an order of magnitude (10 times) larger than the fine-tuned model. This highlights a key observation: without large curated domain-specific training data, general-purpose LLMs struggle to infer chemically meaningful or quantum-chemically grounded properties from SMILES strings.

Together, these results demonstrate that the fine-tuned model not only outperforms large general-purpose LLMs in photoinitiator-relevant quantum-chemical prediction tasks, but also generalizes to unseen molecular distributions effectively. The performance of our fine-tuned model highlights the critical role of high-quality quantum-chemical datasets in achieving accurate molecular property prediction.


\section{Conclusion}

In this work, we present QuantumChem-200K, the first large-scale, openly available molecular dataset focused on photoinitiator-relevant quantum and photochemical properties. By integrating DFT, semi-empirical quantum chemistry, and neural-network predictors in a unified automated workflow, we compute 11 mechanistically meaningful descriptors for more than 200,000 organic molecules. These properties—including TPA cross sections, ISC energies, toxicity, synthetic accessibility, and key physicochemical features—form a comprehensive data foundation for photodissociation modeling, TPP material discovery, and quantum-aware molecular design. 

We further demonstrate that fine-tuning a large language model on domain-specific quantum-chemical data enables substantial improvements in molecular property prediction. The fine-tuned AI assistant achieves competitive or superior performance relative to state-of-the-art LLMs when tested on 3000 previously unseen molecules drawn from VQM24 and ZINC20, with particularly strong accuracy on TPA and excited-state properties critical for photoinitiator evaluation. These results highlight the promise of LLMs as flexible, high-throughput predictors for materials chemistry when equipped with large, high-quality datasets.
QuantumChem-200K, together with the fine-tuned AI assistant and evaluation benchmarks, provides a scalable platform for accelerating photoinitiator design, guiding high-throughput computational screening, and enabling autonomous, agent-based molecular discovery workflows. This work lays the groundwork for future extensions that incorporate excited-state dynamics, time-dependent reactivity, and multi-modal data integration, ultimately moving toward fully automated discovery pipelines for advanced photoresists and photochemical materials.

\section{Data Availability}
The data produced and used by this work is publicly available at: https://huggingface.co/YinqiZeng704. Correspondence for other supporting material should be addressed to Renjie Li. 

\section{Acknowledgement}
This research was supported by the IMB-Illinois Discovery Accelerator Institute (IIDAI). The authors thank Dr. Lynford Goddard, Dr. Paul Braun, Dr. Deming Chen, Dr. Andre Schleife, Noni Ledford, and Dr. Sudhir Gowda for their generous funding support and helpful mentorship. We thank the engineers at the Illinois Computes Campus Cluster for their HPC computing support.

\bibliography{aaai2026.bib}

\newpage
\section{Appendix}
\subsection{Photo-dissociation quantum yield}

Photo-dissociation quantum yield is the most direct and informative indicator of photoinitiator performance, as it measures the fraction of excited molecules that successfully undergo bond cleavage after absorbing photons of a given frequency. Formally, the quantum yield is defined as the rate of the desired dissociation pathway divided by the total rate of all competing photophysical and photochemical processes \cite{Braslavsky2007IUPACGlossary, Budyka2008AromaticAzides}. Traditionally, determining quantum yield relies mostly on experimental measurements: researchers must design a controlled irradiation setup, choose an appropriate resin system, photodissociate the sample, and then infer reaction rates from the resulting polymerization or residual monomer content \cite{Mauri2025Photoreactivity}.

\begin{equation}
  \Phi_{\mathrm{diss}} = 
  \frac{k_{\mathrm{diss}}}{
        k_{\mathrm{diss}} + k_f + k_{\mathrm{IC}} + k_{\mathrm{ISC}}
        + \sum k_{\mathrm{other}}}
  \label{eq:phi_general}
\end{equation}

Another critical quantity governing photodissociation efficiency is the bond dissociation energy (BDE) of the cleavable bond in the excited state. After a molecule absorbs one or two photons and reaches an excited electronic state, it can either relax radiatively through fluorescence or undergo bond cleavage. For dissociation to occur, the absorbed energy must exceed the BDE of the initiating bond, which is why many photoinitiators contain relatively weak, photolabile linkages \cite{blanksby2003bde, morse2019predissociation}. For example, common photoinitiator: DMPA (2,2-dimethoxy-2-phenylacetophenone), undergoes Norrish Type~I $\alpha$-cleavage at the carbonyl--$\alpha$-carbon bond to produce a benzoyl radical and an $\alpha$-alkoxy radical. The BDE for a given cleavage pathway is computed as
\begin{equation}
\mathrm{BDE} = H^\circ(\mathrm{rad}_1) + H^\circ(\mathrm{rad}_2) - H^\circ(\mathrm{parent})
\end{equation}
where $H$ denotes the electronic enthalpy of each species \cite{farmer2024bde}. Lower BDE values generally correlate with more efficient photodissociation and serve as a mechanistic indicator of promising photoinitiator candidates.

\end{document}